# Improving Clinical Target Volume Segmentation Accuracy using Anatomical Priors and Active Learning for the AGITG TOPGEAR Clinical Trial

## Authors

Phillip Chlap (1,2,3)*
Mark Lee (1,3)
Trevor Leong (4,5)
Matthew Field (1,2,3)
Jason Dowling (1,2,6)
Hang Min (1,2,6)
Julie Chu (4,6)
Jennifer Tan (4)
Phillip K. Tran (4)
Tomas Kron (4,5)
Annette Haworth (7)
Martin A. Ebert (8,9,10)
Shalini K. Vinod (1,2,3)
Lois Holloway (1,2,3,7)

1. University of New South Wales, South Western Sydney Clinical School, Sydney, Australia
2. Ingham Institute for Applied Medical Research, Sydney, Australia
3. Liverpool and Macarthur Cancer Therapy Centres, Department of Radiation Oncology, Sydney, Australia
4. Peter MacCallum Cancer Centre, Melbourne, Australia
5. The Sir Peter MacCallum Department of Oncology, University of Melbourne, Melbourne, Australia
6. CSIRO Australian e-Health Research Centre, Herston, Australia
7. Institute of Medical Physics, University of Sydney, Sydney, Australia
8. School of Physics, Mathematics, and Computing, The University of Western Australia, Crawley, Australia
9. Department of Radiation Oncology, Sir Charles Gardiner Hospital, Nedlands, Australia
10. School of Medicine and Population Health, University of Wisconsin, Madison, WI, USA

* Currently employed at Radformation Inc., New York, NY, USA

# Abstract

### *Objective*

Training deep learning-based medical image segmentation models is challenging when limited curated datasets are available. For AGITG TOPGEAR, a gastric cancer trial, the Clinical Target Volume (CTV) is complex and defined by multiple anatomical landmarks, making upfront training data preparation difficult for a segmentation model intended to support automated contour QA within the trial. We investigate the use of anatomical priors, derived from surrounding organ segmentations, to provide spatial context and improve segmentation accuracy for the TOPGEAR CTV. We also evaluate active learning, which iteratively expands the training dataset by selecting additional samples expected to improve model performance.

### *Approach*

One hundred TOPGEAR CT scans were retrospectively analyzed. An initial set of 10 expert-contoured cases was used to train an nnU-Net model. TotalSegmentator generated a voxel-wise anatomical prior map from surrounding structures, provided to the model as an additional input channel. Active learning was simulated prospectively over four iterations, selecting cases based on model uncertainty and segmentation performance. All models were trained using five-fold cross-validation to produce an ensemble-based uncertainty measure. Evaluation was performed on a hold-out testing set of 50 cases.

### *Main results*

Inclusion of the anatomical prior improved CTV segmentation accuracy, increasing mean Dice Similarity Coefficient (DSC) from 0.84 to 0.86. Active learning similarly improved performance to 0.86, with the greatest benefit in the final round. The combination of the anatomical prior with active learning achieved the highest accuracy with a DSC of 0.87. Model uncertainty estimation correlated with DSC, supporting its utility in identifying suboptimal predictions and guiding active learning.

### *Significance*

Anatomical priors and active learning each improved CTV segmentation accuracy and generalizability, with their combination achieving the best performance, supporting the integration of these techniques into segmentation model development for automated contour QA in radiotherapy clinical trials.
AGITG AG0407GR/TROG 08.08 "TOPGEAR" (ClinicalTrials.gov: NCT01924819).

## Introduction

Deep learning-based segmentation requires large, curated datasets for training (Sahiner *et al.*, 2019). In medical imaging, obtaining and curating such datasets is challenging and resource intensive (Tajbakhsh *et al.*, 2020). Greater diversity in patient anatomy and imaging characteristics improves model generalizability by reducing epistemic uncertainty (Abdar *et al.*, 2021).

We investigate training a segmentation model for the TOPGEAR trial's Clinical Target Volume (CTV) to support automated contour QA, using a small initial dataset and techniques to improve segmentation accuracy. In this application, the automatically generated segmentation is intended to serve as a reference against which submitted contours can be compared, flagging cases where the manual CTV deviates from the model prediction for further review. AGITG AG0407GR/TROG 08.08 "TOPGEAR" (ClinicalTrials.gov: NCT01924819) compared perioperative chemotherapy alone with or without preoperative chemoradiotherapy for patients with resectable gastric cancer (Leong *et al.*, 2015, 2017, 2024). Its CTV was defined by anatomical landmarks throughout the abdomen with explicit structures to be included or excluded from the volume. Quality assurance (QA) was performed on sampled patients' treatment plans to ensure CTV conformance to protocol (Lukovic *et al.*, 2023). Such QA processes are recommended in modern radiotherapy trials to maintain trial integrity and minimize negative impact on patient outcomes (Peters *et al.*, 2010; Mir *et al.*, 2020).

Radiotherapy trials introducing novel structures, such as the TOPGEAR CTV, make curation of large pre-trial datasets challenging, as representative imaging and gold-standard contours for training are required. Models trained on small datasets often perform sub-optimally due to limited anatomical and imaging diversity (Chlap *et al.*, 2024). Although data augmentation can improve performance, its ability to extend generalization beyond the anatomical or imaging characteristics already present in the original training data remains limited (Chlap *et al.*, 2021).

Pre-trained anatomical segmentation models can offer valuable contextual information, especially when limited training data is available. One such model is TotalSegmentator, which segments 117 anatomical structures on CT scans with high accuracy (Wasserthal *et al.*, 2023). Although it does not delineate the TOPGEAR CTV directly, TotalSegmentator segments several key structures that define or border the CTV, such as the pancreas, duodenum, and stomach, which can be used as anatomical priors to provide spatial context during model training. TotalSegmentator is based on nnU-Net, a self-configuring segmentation framework that consistently achieves state-of-the-art performance across diverse datasets (Isensee *et al.*, 2021).

Anatomical priors have been used to improve segmentation performance, particularly for target volumes defined by their spatial relationship to surrounding organs. These priors can be generated using the output of segmentation models to provide voxel-wise spatial context of adjacent structures. This additional context can be especially valuable for cases with high anatomical variability or heterogeneous imaging characteristics. This approach has been explored extensively for mediastinal lymph node segmentation (Feulner *et al.*, 2013; Bouget *et al.*, 2023; Engelson *et al.*, 2024; Mathai, Liu and Summers, 2024).

Active learning is a strategy for improving model performance by efficiently increasing the training dataset size (Budd, Robinson and Kainz, 2021). It involves iteratively retraining the model using additional data selected based on its potential to improve model accuracy. Selection is typically guided by the previous model's predictions, with preference given to cases with high uncertainty, often those with atypical anatomy or imaging characteristics not well represented in the training set (Wang *et al.*, 2024). Furthermore, active learning can help model generalization by avoiding overfitting to overrepresented samples when faced with an imbalanced dataset (Liu *et al.*, 2023).

In this study, we evaluate the use of anatomical priors and active learning to improve segmentation of the TOPGEAR CTV. Although the trial is now complete, we simulate a prospective setting in which cases become available sequentially. Within each round, only a subset of cases was selected for retraining based on model confidence and accuracy, reflecting a practical active learning workflow for segmentation model development in future radiotherapy clinical trials. As the reliability of this comparison depends on the accuracy of the automatically generated reference segmentation, improving overall model accuracy is a necessary step toward its future application for automated contour QA.

# Method

## Dataset

The dataset $\mathcal{D}$ consisted of planning CT scans and structure sets for 100 patients selected from 174 TOPGEAR patients sampled for manual QA during the trial.

An initial training set $\mathcal{D}_{train}$ consisted of the first 10 patients treated in the trial. For each case, five radiation oncologists (RO) independently contoured the CTV, denoted by $S^1(x), S^2(x), \ldots, S^5(x)$. A consensus workshop was held to ensure protocol adherence, with contours adjusted as needed (Cloak *et al.*, 2019). The reference segmentation $S^{ref}(x)$ was generated using the STAPLE method (Warfield, Zou and Wells, 2004):

$$S^{\text{ref}}(x) = \text{STAPLE}\left(\{S^{(i)}(x) \mid i = 1, \ldots, 5\}\right)$$

This gold-standard dataset was intended to simulate the annotation effort feasible at trial commencement.

A trial set $\mathcal{D}_{trial}$ of 40 cases and a hold-out testing set $\mathcal{D}_{test}$ of 50 cases were randomly sampled from the remaining pool. The clinical CTV $S^{ref}(x)$ was used as the reference contour for training and evaluation. This reflects the intended contour QA application, in which the model's prediction is compared against the contour submitted by the treating center to identify potential protocol deviations.

The TOPGEAR trial protocol specifies that when tumor is present in the esophagus, the CTV should extend 4cm into this region. This scenario was excluded during preparation of $\mathcal{D}_{train}$ and this region was not contoured by the five experts. To ensure consistency across the datasets, this region was excluded from $S^{\text{ref}}(x)$ in $\mathcal{D}_{trial}$ and $\mathcal{D}_{test}$ by removing connected components overlapping with the esophagus (predicted by TotalSegmentator) that were smaller than $50\text{mm}^2$.

$\mathcal{D}_{trial}$ was used to simulate prospective case availability during the trial (see Active Learning), while $\mathcal{D}_{test}$ was reserved for final evaluation of the segmentation approaches. An overview of the dataset is provided in Table 1. Data conversion and preparation for nnU-Net training were performed using PyDicer (Chlap *et al.*, 2025).

Ethics approval to conduct this work using the TOPGEAR (2019/ETH04489) dataset was obtained via the Human Research Ethics Committee (HREC) at Liverpool hospital (South Western Sydney Local Health District, Australia).

*Table 1: TOPGEAR dataset split used for experiments in Anatomical Prior (AP) and Active Learning (AL) model training. The number of cases selected are shown for the trial set with the total number of cases included in a given round.*

| Dataset | Active Learning Iteration | # Total Cases | # Cases after exclusion | Selected for training (no AP) | Selected for training (with AP) |
|---|---|---|---|---|---|
| **Initial** | | 10 | 10 | 10 | 10 |
| **Trial** | | 40 | 32 | 15 | 12 |
| | $\mathcal{G}_1$ | *10* | *7* | *4 (14)* | *2 (12)* |
| | $\mathcal{G}_2$ | *10* | *7* | *4 (18)* | *2 (14)* |
| | $\mathcal{G}_3$ | *10* | *9* | *5 (23)* | *4 (18)* |
| | $\mathcal{G}_4$ | *10* | *9* | *2 (25)* | *4 (22)* |
| **Hold-out** | | 50 | 49 | - | - |
| **Total** | | 100 | 91 | 25 | 22 |

## Model training

The nnU-Net (v2.5.1) 3D full-resolution configuration was used to train all TOPGEAR CTV segmentation models, as preliminary testing showed it outperformed the 2D and low-resolution 3D variants. Each model was trained using five-fold cross-validation, with 80% of data for training and 20% for validation. The final model ensemble was defined as:

$$\mathcal{M} = \{\mathcal{F}_1, \mathcal{F}_2, \mathcal{F}_3, \mathcal{F}_4, \mathcal{F}_5\}$$

where each $\mathcal{F}_i$ is the best model from its fold, selected based on validation performance through the 1000 epochs trained.

Each model $\mathcal{F}_i \in \mathcal{M}$ produces a voxel-wise softmax probability map $P_i(x) \in [0,1]^{H\times W\times D\times C}$, where $C = 2$ corresponds to foreground and background classes used for the CTV model. The final ensemble prediction is the mean across all folds:

$$P_\mu(x) = \frac{1}{5}\sum_{i=1}^{5} P_i(x)$$

The final segmentation $\hat{S}(x)$ is computed by assigning each voxel the class with the maximum softmax probability:

$$\hat{S}(x) = \arg\max_{c\in C} P_\mu\,(x)_c$$

Uncertainty was estimated from the voxel-wise variability across ensemble predictions. This estimate reflects the model's confidence and can identify cases likely to underperform. A boundary region $\mathcal{B}_\delta(x)$ is defined as the set of voxels within a distance $\delta$ mm from the final prediction.

$$\mathcal{B}_\delta(x) = \{v \in \text{voxels} \mid |D(v)| < \delta\}$$

Here, $D(v)$ is the signed distance from voxel $v$ to the predicted contour. For each voxel $v$, the standard deviation across the softmax predictions from all folds is computed:

$$\sigma(v) = \sqrt{\frac{1}{|\mathcal{M}|}\sum_{i=1}^{5}(P_i(x)v - P(x)_v)^2}$$

A voxel is considered uncertain if $\sigma(v)$ is above a threshold $\lambda$. An uncertainty score is calculated as the proportion of uncertain voxels within $\mathcal{B}_\delta(x)$:

$$U(x) = \frac{|\{v \in \mathcal{B}_\delta(x) \mid \sigma(v) > \lambda\}|}{|\mathcal{B}_\delta(x)|}$$

Model confidence is then defined as:

$$\mathcal{C}(x) = 1 - U(x)$$

In this study, $\delta$ = 5mm and $\lambda$ = 0.2 were chosen empirically. The initial model $\mathcal{M}_0$ was trained on $\mathcal{D}_{train}$ using this approach.

All training and experimentation were performed on an Ubuntu 22.04 system with 16 CPU Cores, 128 GB RAM and an NVIDIA Quadro 8000 GPU with 48 GB RAM.

## Anatomical Prior

The TOPGEAR trial defines the CTV based on anatomical landmarks, including the esophagus, stomach, duodenum, hepatogastric ligament, porta hepatis, pancreas, and major vessels such as the celiac and superior mesenteric arteries. For example, the CTV is described to extend inferiorly down to the 3rd part of the duodenum. Additional structures, such as the liver, kidneys, heart, and lungs, guide the CTV position and are considered organs at risk (OAR) to which dose should be limited.

TotalSegmentator is a publicly available model trained on over 1,200 CT and MR scans to segment 117 anatomical structures (Wasserthal *et al.*, 2023). Many of these structures correspond to those used in TOPGEAR for CTV definition or OAR constraints. While clinician-provided OAR segmentations were available, they may contain protocol deviations themselves or inter-observer variability and were not uniformly available across the dataset. In contrast, TotalSegmentator offers consistent, high-quality structure definitions across all cases, making it better suited for use as an anatomical prior during model training. We hypothesized that including these standardized segmentation masks alongside the CT image would provide spatial context and improve CTV segmentation accuracy.

A voxel-wise anatomical label map $S_{\mathrm{TS}}(x) \in \{0,1,\ldots,117\}^{H\times W\times D}$ was generated using TotalSegmentator (v2.4.0), assigning each voxel to one of the 117 structures or background (0). Since the structures do not overlap, a single-channel map was sufficient. The map was normalized to the range $[0,1]$ to form the anatomical prior:

$$AP(x) = \frac{S_{\mathrm{TS}}(x)}{117},$$

where $AP(x) \in [0,1]^{H\times W\times D}$ represents the anatomical prior. $AP(x)$ was concatenated with the CT image $x$, forming a two-channel input $x' = \mathrm{concat}\big(x, AP(x)\big) \in R^{2\times H\times W\times D}$.

To evaluate this approach, we trained a model $\mathcal{M}_{0AP}$ on $\mathcal{D}_{train}$ using $x'$ as input and compared its performance to $\mathcal{M}_0$, trained without anatomical priors.

## Active learning

To evaluate the benefit of increasing dataset size during a prospective trial, we simulated this setting treating cases in $\mathcal{D}_{trial}$ sequentially. These cases were processed in $K$ rounds, in groups of 10 cases each $\mathcal{G}_{k}$. This setup assumes an initial model $\mathcal{M}_0$ is available, trained with an initial dataset $\mathcal{D}_{train}$. Two expert radiation oncologists reviewed each case to ensure that the reference contour $S^{ref}(x)$ conformed to the trial protocol. Contours were corrected if necessary or excluded if image quality was inadequate for accurate CTV definition.

During each round $k$, a model $\mathcal{M}_{k-1}$ was applied to all cases $x \in \mathcal{G}_{k}$, producing:

- A predicted segmentation $\hat{S}(x)$,
- A confidence score $\mathcal{C}(x)$,
- The Dice Similarity Coefficient (DSC) between $S^{ref}(x)$ and $\hat{S}(x)$.

Cases were selected if they were flagged as low confidence or low DSC:

$$select(x) = \begin{cases} True\ if\ \mathcal{C}(x) < P_{\rho}(\mathcal{G}_{k}) \\ True\ if\ DSC(S^{ref}(x), \hat{S}(x)) < \gamma \\ False\ otherwise \end{cases}$$

The selected cases formed the updated set:

$$\mathcal{R}_{k} = \{x \in \mathcal{G}_{k} \mid select(x)\}$$

The training set was updated accordingly:

$$\mathcal{D}_{train}{}^{(k)} = \mathcal{D}_{train}{}^{(k-1)} \cup \mathcal{R}_{k}$$

New models were trained using the updated set, both with ($\mathcal{M}_{kAP}$) and without ($\mathcal{M}_{k}$) anatomical priors. Each new model was trained from scratch on the full updated training set, rather than by continuing training from the previous round's model checkpoint.

Active learning rounds were performed for $k = 1, \ldots, K$, with $K = 4$ based on dataset size and RO review capacity. Selection parameters were set to $\rho = 25$ (confidence percentile) and $\gamma = 0.85$ (DSC threshold) and were chosen empirically. This selection-based approach reflects a practical active learning workflow that prioritizes cases expected to improve model performance.

### Model Evaluation

Ten models were evaluated across the anatomical prior and active learning approaches: $\{\mathcal{M}_0, \mathcal{M}_1, \mathcal{M}_2, \mathcal{M}_3, \mathcal{M}_4, \mathcal{M}_{0AP}, \mathcal{M}_{1AP}, \mathcal{M}_{2AP}, \mathcal{M}_{3AP}, \mathcal{M}_{4AP}\}$. Each model was applied to $\mathcal{D}_{test}$ and evaluated using four metrics: Dice Similarity Coefficient (DSC) (Sorensen, 1948), Surface DSC (sDSC, $\tau = 3mm$) (Nikolov *et al.*, 2018), Hausdorff Distance (HD) (Dubuisson and Jain, 1994) and Mean Surface Distance (MSD). These overlap and boundary-based metrics were selected for their suitability in assessing TOPGEAR CTV segmentation (Maier-Hein *et al.*, 2024). The Pearson correlation coefficient between $\mathcal{C}(x)$ and DSC was used to assess the model confidence score. All similarity metrics were computed using PlatiPy (Chlap and Finnegan, 2023).

## Results

Including the anatomical prior map improved segmentation performance on $\mathcal{D}_{test}$. As shown in Figure 1, $\mathcal{M}_{0AP}$ outperformed $\mathcal{M}_0$ for all metrics, with mean DSC increasing from 0.84±0.09 to 0.86±0.05 with reduced variability. A paired t-test showed statistically significant improvement ($p<0.05$) across all metrics.

Active learning without anatomical priors resulted in incremental improvements, with increased DSC from $\mathcal{M}_0$ to $\mathcal{M}_4$ reaching 0.86±0.05. Improvement in $\mathcal{M}_1, \mathcal{M}_2$ and $\mathcal{M}_3$ were minimal and not statistically significant (Figure 2). When anatomical priors were included, active learning had a smaller effect. However, $\mathcal{M}_{4AP}$ achieved the best overall performance with DSC of 0.87±0.03. $\mathcal{M}_{0AP}$ and $\mathcal{M}_4$ both achieved a mean DSC of 0.86 indicating that all 4 rounds of active learning were required before equivalent performance of using only anatomical priors was achieved.

CTV inter-observer variability (IoV) was assessed on $\mathcal{D}_{train}$ using the five expert segmentations $S^1(x), S^2(x), \ldots, S^5(x)$. IoV metric results and as well as model evaluation metrics are presented in Table 2.

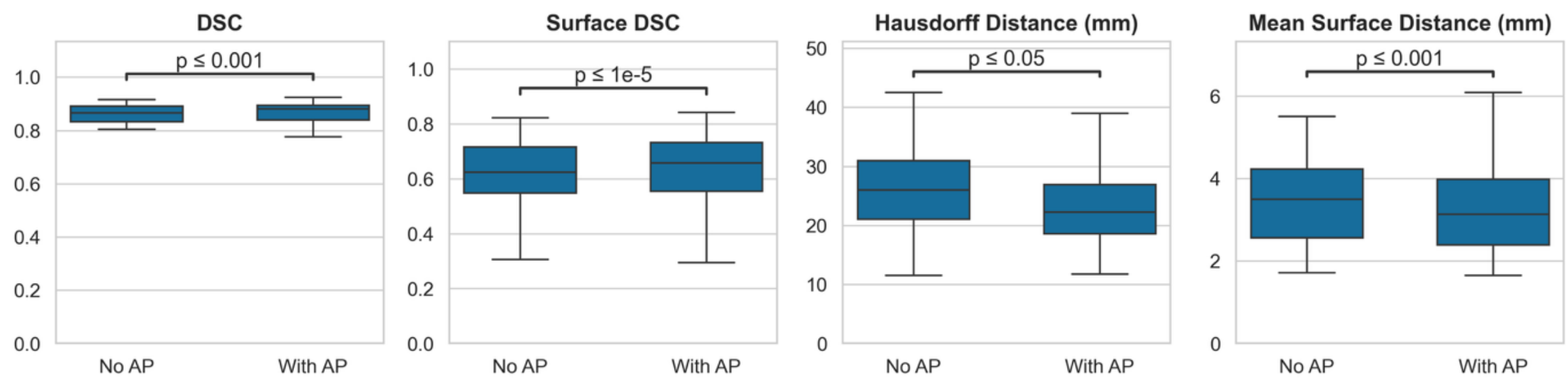

*Figure 1: Segmentation results comparing the initial model trained with and without anatomical priors when applied to $\mathcal{D}_{test}$. A paired T-Test was used to show significance between $\mathcal{M}_0$ and $\mathcal{M}_{0AP}$.*

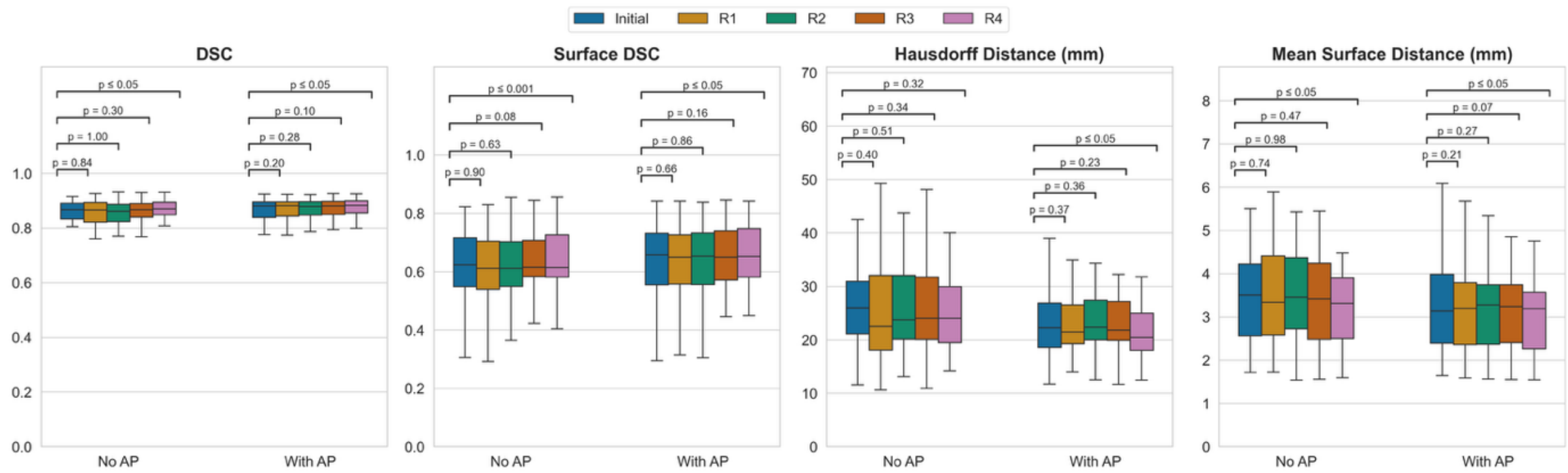

*Figure 2: Segmentation results comparing models across active learning iterations applied to $\mathcal{D}_{test}$. Paired T-Test was used to show the statistical significance between model $\mathcal{M}_{1-4}$ and $\mathcal{M}_0$.*

Figure 3 shows the correlation between the confidence score $\mathcal{C}(x)$ and DSC with a consistent positive Pearson correlation across models. This supports the utility of $\mathcal{C}$ for active learning and future integration of the model in trial quality assurance.

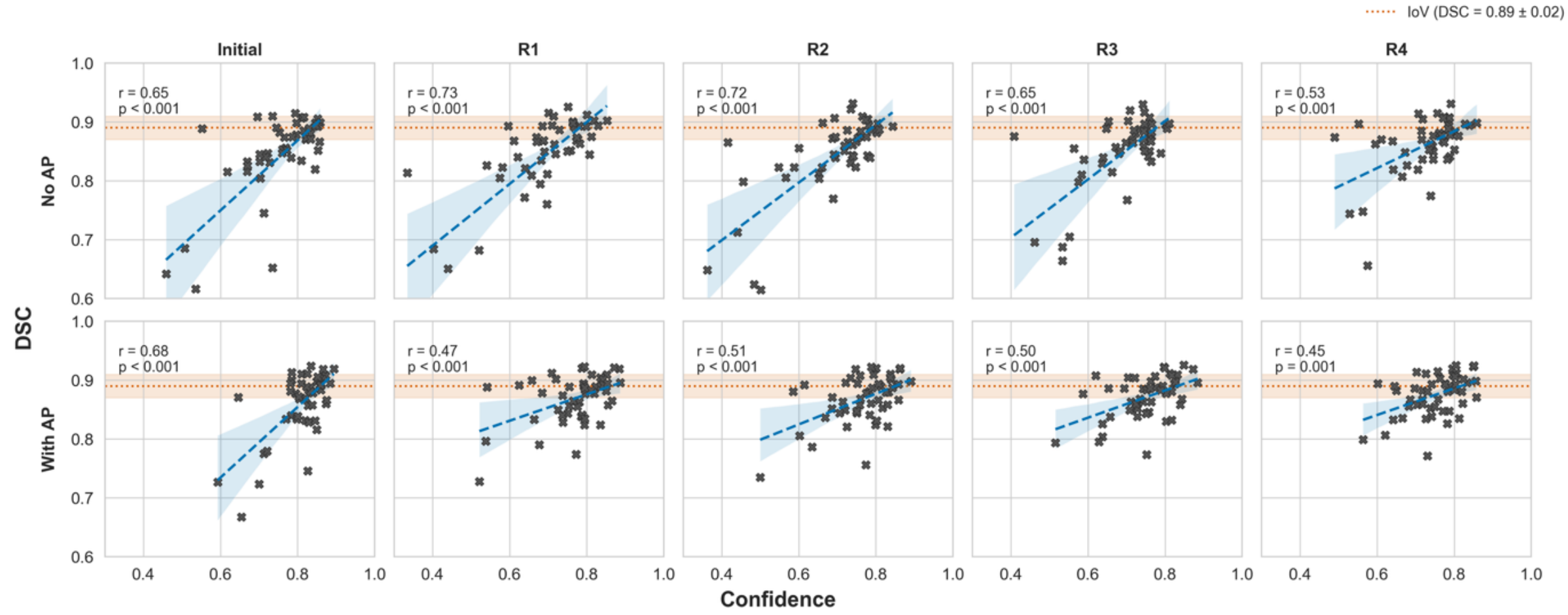

*Figure 3: Correlation between confidence score $\mathcal{C}$ and the DSC for each model trained when applied to $\mathcal{D}_{test}$. Correlation was assessed using the Pearson Correlation Coefficient. The shaded band and dotted line indicate the inter-observer variability (IoV) in DSC (0.89 ± 0.02), representing an approximate upper bound for model performance.*

Approximate average CTV model training times for a single fold was 22 hours when anatomical priors were included and 18 hours without. Approximate inference time for CTV segmentation models with anatomical priors was 7 minutes per image which was reduced to 6 minutes without. TotalSegmentator inference runtime to generate anatomical prior maps was approximately 6 minutes per image.

*Table 2: Results for each of the models trained with and without Anatomical Prior (AP) across each of the Active Learning (AL) iterations evaluated on the hold-out dataset.*

| | DSC | | sDSC | | HD (mm) | | MSD (mm) | |
|---|---|---|---|---|---|---|---|---|
| **IoV** | 0.89 ± 0.02 | | 0.74 ± 0.06 | | 20 ± 3.9 | | 2.35 ± 1.43 | |
| **Dataset** | **No AP** | **With AP** | **No AP** | **With AP** | **No AP** | **With AP** | **No AP** | **With AP** |
| **Initial** | 0.84 ± 0.09 | 0.86 ± 0.06 | 0.60 ± 0.14 | 0.64 ± 0.13 | 32.9 ± 26.7 | 24.9 ± 11.0 | 4.31 ± 2.82 | 3.46 ± 1.60 |
| $\mathcal{R}_1$ | 0.84 ± 0.08 | 0.87 ± 0.04 | 0.60 ± 0.14 | 0.64 ± 0.12 | 29.5 ± 18.8 | 24.1 ± 9.4 | 4.42 ± 3.24 | 3.34 ± 1.35 |
| $\mathcal{R}_2$ | 0.84 ± 0.08 | 0.87 ± 0.04 | 0.61 ± 0.13 | 0.64 ± 0.12 | 30.3 ± 18.6 | 26.5 ± 13.3 | 4.32 ± 3.18 | 3.35 ± 1.36 |
| $\mathcal{R}_3$ | 0.85 ± 0.07 | 0.87 ± 0.04 | 0.62 ± 0.12 | 0.65 ± 0.12 | 29.3 ± 16.7 | 23.8 ± 9.1 | 4.10 ± 2.80 | 3.27 ± 1.21 |
| $\mathcal{R}_4$ | 0.86 ± 0.05 | **0.87 ± 0.03** | 0.63 ± 0.12 | **0.65 ± 0.12** | 28.8 ± 18.3 | **23.0 ± 9.9** | 3.65 ± 1.91 | **3.20 ± 1.22** |

## Discussion

Training a segmentation model for the TOPGEAR CTV posed both unique challenges and opportunities. The dataset comprised imaging from multiple international centers with variation in scanner types, acquisition protocols, and the use of both intravenous and oral contrast. This heterogeneity, while representative of real-world clinical trial settings, introduces significant variability in anatomical appearance and imaging characteristics. To address these challenges, the best-performing model, $\mathcal{M}_{4AP}$ (DSC = 0.87), incorporated both anatomical priors and active learning to enable the model to improve generalization across the dataset. However, its performance was comparable to using either technique independently. This suggests that both approaches provide similar benefits by introducing spatial context or anatomical diversity to the training set, and that their combination yields diminishing returns. This performance approached the inter-observer variability (IoV) observed on $\mathcal{D}_{train}$ (DSC = 0.89), which likely represents an upper bound for model performance on this task, though it was measured on a limited and potentially less diverse subset of the full dataset (Jungo *et al.*, 2018).

The model's confidence score $\mathcal{C}(x)$ was well-correlated with DSC, confirming its utility in identifying low-performing cases. This highlights its potential value in active learning, where low-confidence predictions may reflect either underrepresented anatomical diversity or limitations in the input data, such as poor image quality or artefacts. Beyond training, confidence scores could support deployment by alerting users to predictions that are potentially unreliable and may warrant closer review or manual refinement. In the context of automated contour QA, low-confidence cases could be referred directly for manual review.

The anatomical prior was designed to provide spatial context based on surrounding anatomy. For example, the TOPGEAR protocol specifies that the CTV should extend inferiorly to include the third part of the duodenum, which was frequently missed during the trial as uncovered during QA (Lukovic *et al.*, 2023). Since the prior map includes a prediction of the duodenum, models using the prior map better covered this landmark. Figure 4 illustrates this improvement, where $\mathcal{M}_{0AP}$ produced more accurate inferior coverage than $\mathcal{M}_0$ on a contrast enhanced scan. No contrast-enhanced scans were present in $\mathcal{D}_{train}$, suggesting that the anatomical prior helped the model generalize to imaging characteristics not seen during training. The anatomical prior also improved segmentation in the superior region around the stomach. While the contrast-enhanced appearance likely contributed to reduced accuracy in $\mathcal{M}_0$, the anatomical prior provided explicit localization of the stomach, enabling more accurate CTV coverage in this region by $\mathcal{M}_{0AP}$.

These findings are consistent with previous work in mediastinal lymph node segmentation, where anatomical priors improved accuracy by providing spatial context. Feulner et. al. used segmentations of the heart chambers and esophagus to guide their model (Feulner *et al.*, 2013). Engelson et. al. evaluated various anatomical prior generation methods, including the use of TotalSegmentator, but found no significant benefit (Engelson *et al.*, 2024). In contrast, Mathai et. al. applied a similar TotalSegmentator-based strategy and reported improved performance (Mathai, Liu and Summers, 2024). Bouget et. al. used vascular structures, including the azygos vein, subclavian arteries and brachiocephalic veins, as priors and suggested that inclusion of additional organs was needed for optimal benefit (Bouget *et al.*, 2023). While TotalSegmentator provides high quality segmentations, it was not developed specifically for radiotherapy. Task-specific models developed for radiotherapy planning, whether open-source or commercial, may offer better suitability to guide radiotherapy planning structure segmentation (Erdur *et al.*, 2024). These alternatives could also be explored as sources of anatomical priors in future work (Doolan *et al.*, 2023; Finnegan *et al.*, 2023; Constantinou *et al.*, 2025).

The anatomical prior is particularly well suited to the TOPGEAR CTV, which is anatomically defined. Priors are most beneficial when:

1. The training dataset is small and underrepresents anatomical variability,
2. The structure is defined relative to nearby anatomy,
3. The structure boundary is not aligned with strong intensity gradients.

Several radiotherapy target volumes meet these criteria. For example, a prostate bed CTV is defined in relation to organs like the bladder and rectum (Sidhom *et al.*, 2008). Here the use of distance maps to nearby structures may provide even better spatial context than a voxel-wise label maps. Post-operative breast tumor bed segmentation has been shown to improve with the addition of pre-surgery prior information (Huang *et al.*, 2024). Anatomical prior information has also been used to guide brain tumor segmentation training where these were applied to masking inputs for vision transformers (Wang *et al.*, 2023).

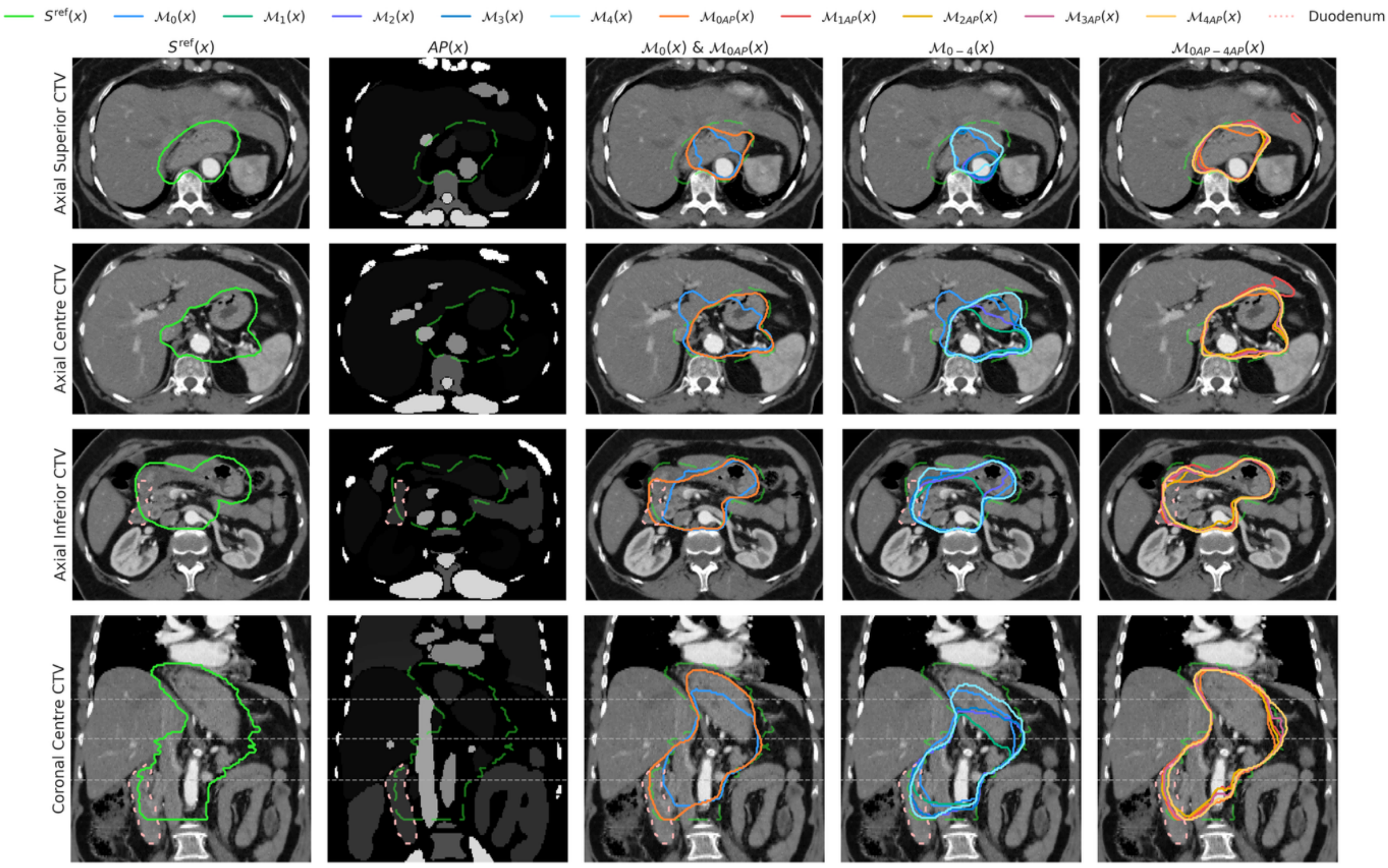

*Figure 4: Example case depicting the manual CTV segmentation, the anatomical prior map, initial models with and without AP, and active learning model results without AP and with AP. The Duodenum is shown (dotted contour) as an example landmark. Rows 1-3 show axial slices from the superior, center and inferior region respectively. The bottom row shows the coronal slice with the axial slice location indicated (dashed grey line).*

Active learning also aims to increase anatomical diversity in the training set by selectively introducing cases based on model performance and confidence. This helps mitigate overfitting to early training examples. Nath et al. applied active learning to reduce the annotation burden for segmentation of the hippocampus on MRI and the pancreas on CT (Nath *et al.*, 2021). Model uncertainty estimation was used to iteratively select cases from a larger unlabeled pool. With this approach, models trained with 25% of the hippocampus data and 50% of the pancreas data achieved equivalent performance to models trained on the full dataset. Similarly, Mahapatra et al. showed that lung segmentation on chest X-rays achieved equivalent accuracy using only 35% of the data when selected through active learning (Mahapatra *et al.*, 2018). For brain tumor segmentation on MR, Kim et. al. demonstrated that uncertainty-based selection outperformed random sampling, achieving comparable performance with just 35% of the data (Kim *et al.*, 2024).

In a prospective clinical trial setting, training data is not available upfront but accumulates as the trial progresses. Men et. al. used model uncertainty to guide case selection for OAR segmentation in a lung cancer trial, improving segmentation model performance through active learning for use in automating contour quality assurance (Men *et al.*, 2020). Fowler et. al. explored active learning in this setting by sequentially sampling data during two ophthalmology trials for eye disease detection using optical coherence tomography imaging (Fowler *et al.*, 2023). They found that sampling sequentially allowed their diagnostic model to adapt to evolving trial data.

In our study, we applied active learning using a hybrid selection strategy combining model confidence and DSC to guide inclusion of additional training data. With this approach, $\mathcal{M}_4$ can be seen to better cover the duodenum and stomach relative to $\mathcal{M}_0$ in Figure 4, though intermediate models like $\mathcal{M}_1$ performed worse in some regions. This likely reflects case variability and the limitations of sequential data availability, which mimics a prospective trial scenario. This is seen in the metrics evaluated across all cases, where improvement was not significant until the final round.

Given that $\mathcal{M}_4$ produced the most gain in accuracy, additional rounds of active learning may further benefit model performance. In typical AL workflows, rounds are continued until performance plateaus (Ren *et al.*, 2022). In a clinical trial setting, AL could extend throughout the study if further benefit to the model was expected.

While gastric CTV segmentation is relatively novel, it has been previously explored. Previous work used a probabilistic U-Net to capture uncertainty, achieving a DSC of 0.67 on the same $\mathcal{D}_{train}$ dataset (Chlap *et al.*, 2024). Xu et. al. reported a stomach cancer CTV model with a DSC of 0.74 using 150 training cases (Xu *et al.*, 2021), highlighting the effectiveness of the methods used in this study despite limited data.

Determining when a segmentation model is “good enough” depends on its intended use. When applied during a trial, the model can support QA by flagging manual CTVs that deviate from the model prediction. While high accuracy is desirable, even partial performance can be valuable if errors lead to false positives rather than missed violations. The model may be used after trial completion by clinics aiming to treat to the trial protocol. The improvements in DSC achieved through anatomical priors and active learning in this study (0.84 to 0.87) are expected to reduce both missed protocol deviations and unnecessary case flagging when the model is applied for automated contour QA in future work. Evaluation of the resulting contour QA performance itself, including its sensitivity and specificity for detecting protocol deviations, will be addressed in forthcoming work focused specifically on this application.

Our model was trained to predict a single “correct” CTV. However, through IoV we know that aleatoric uncertainty exists. The five expert segmentations in $\mathcal{D}_{train}$ were refined before being combined using STAPLE. While clinical contours in $\mathcal{D}_{trial}$ and $\mathcal{D}_{test}$ were QA’d, they may not reflect exactly the same distribution. Thus, achieving the IoV upper limit on these datasets may not be feasible.

Several parameters in our uncertainty estimation and active learning setup, including $\delta$, $\lambda$, $\rho$ and $\gamma$, were empirically chosen. Future implementations would need to tune these based on the dataset, anatomical site, and the structure itself. Similarly, benchmarking against alternative organ-guided segmentation or active-learning selection methods was considered outside the scope of this study, which prioritized establishing a practically deployable approach for automated contour QA within the TOPGEAR trial.

Finally, we recommend that future trials aiming to build segmentation models prospectively incorporate anatomical priors when appropriate. These provide valuable information from a broader anatomical distribution, which may not be captured through active learning alone. Implementing real-time active learning during a trial can be logistically challenging, requiring infrastructure for data curation and retraining. However, if automated segmentation is used during a trial, model performance should be monitored, and a retraining strategy should be defined if deemed necessary (Chlap *et al.*, 2026).

## Conclusion

Both anatomical priors and active learning improved CTV segmentation accuracy for TOPGEAR. Anatomical priors enabled strong performance from a small, curated initial dataset by providing spatial context from surrounding anatomy. Active learning further refined model generalization by selectively incorporating additional cases based on model confidence and segmentation accuracy. Combined, these techniques achieved the highest overall accuracy. These findings support their use in model development for automated contour QA in prospective radiotherapy clinical trials.

## Acknowledgements

The TOPGEAR trial was funded through the following bodies: Cancer Australia, Cancer Council Australia, Canadian Institutes of Health Research (CIHR), the European Organisation for Research and Treatment of Cancer (EORTC) Academic Research Fund, and the NHMRC. The Australasian Gastro-Intestinal Trials Group (AGITG) was the Sponsor of the TOPGEAR trial, which was coordinated through the National Health and Medical Research Council (NHMRC) Clinical Trials Centre, University of Sydney. The trial incorporated a comprehensive RTQA program coordinated by the Trans-Tasman Radiation Oncology Group (TROG). TOPGEAR was conducted in collaboration with the Canadian Cancer Trials Group (CCTG) and the European Organisation for Research and Treatment of Cancer (EORTC).

This work was additionally supported by NHMRC grant APP1102198. High Performance Computing infrastructure was provided through an award from the Centre for Oncology Education and Research Translation (CONCERT). P.C. is supported by an Australian Government Research Training Program (RTP) Fee Offset Scholarship.

The authors would like to thank TROG and AGITG for supporting this work and making the TOPGEAR dataset available for secondary analysis. We also thank Prof. Laurence Court (MD Anderson Cancer Center) for his valuable insights throughout this study.

## Declaration of Generative AI and AI-assisted technologies in the writing process

Statement: During the preparation of this work the primary author used OpenAI's ChatGPT to assist with editing and improving the clarity of the manuscript. After using this tool, the authors reviewed and edited the content as needed and take full responsibility for the content of the publication.

## References

- Abdar, M. *et al.* (2021) 'A review of uncertainty quantification in deep learning: Techniques, applications and challenges', *Information Fusion*, 76, pp. 243–297. doi:10.1016/j.inffus.2021.05.008.
- Bouget, D. *et al.* (2023) 'Mediastinal lymph nodes segmentation using 3D convolutional neural network ensembles and anatomical priors guiding', *Computer Methods in Biomechanics and Biomedical Engineering: Imaging and Visualization*, 11(1), pp. 44–58. doi:10.1080/21681163.2022.2043778.
- Budd, S., Robinson, E.C. and Kainz, B. (2021) 'A survey on active learning and human-in-the-loop deep learning for medical image analysis', *Medical Image Analysis*. Elsevier B.V., p. 102062. doi:10.1016/j.media.2021.102062.

- Chlap, P. *et al.* (2021) ‘A review of medical image data augmentation techniques for deep learning applications’, *Journal of Medical Imaging and Radiation Oncology*, pp. 1–19. doi:10.1111/1754-9485.13261.
- Chlap, P. *et al.* (2024) ‘Uncertainty estimation using a 3D probabilistic U-Net for segmentation with small radiotherapy clinical trial datasets’, *Computerized Medical Imaging and Graphics*, 116, p. 102403. doi:10.1016/j.compmedimag.2024.102403.
- Chlap, P. *et al.* (2025) ‘PyDicer: An open-source python library for conversion and analysis of radiotherapy DICOM data’, *SoftwareX*, 29, p. 102010. doi:10.1016/j.softx.2024.102010.
- Chlap, P. *et al.* (2026) ‘Implementation of an automated contour quality assurance tool within the TROG 18.01 NINJA trial’, *Radiotherapy and Oncology*, 214, p. 111269. doi:10.1016/j.radonc.2025.111269.
- Chlap, P. and Finnegan, R.N. (2023) ‘PlatiPy: Processing Library and Analysis Toolkit for Medical Imaging in Python’, *Journal of Open Source Software*, 8(86), p. 5374. doi:10.21105/joss.05374.
- Cloak, K. *et al.* (2019) ‘OC-049: Avoiding garbage in: A consensus workshop for refining gastric cancer radiotherapy atlas data’, in *Radiotherapy and Oncology*. Elsevier, pp. 141 (S20-S21).
- Constantinou, A.D. *et al.* (2025) ‘OSAIRIS: Lessons Learned From the Hospital-Based Implementation and Evaluation of an Open-Source Deep-Learning Model for Radiotherapy Image Segmentation’, *Clinical Oncology*, 37. doi:10.1016/j.clon.2024.10.032.
- Doolan, P.J. *et al.* (2023) ‘A clinical evaluation of the performance of five commercial artificial intelligence contouring systems for radiotherapy’, *Frontiers in Oncology*, 13. doi:10.3389/fonc.2023.1213068.
- Dubuisson, M.P. and Jain, A.K. (1994) ‘A modified Hausdorff distance for object matching’, in *Proceedings - International Conference on Pattern Recognition*, pp. 566–568. doi:10.1109/ICPR.1994.576361.
- Engelson, S. *et al.* (2024) ‘Comparison of anatomical priors for learning-based neural network guidance for mediastinal lymph node segmentation’, in. SPIE-Intl Soc Optical Eng, p. 53. doi:10.1117/12.3006158.
- Erdur, A.C. *et al.* (2024) ‘Deep learning for autosegmentation for radiotherapy treatment planning: State-of-the-art and novel perspectives’, *Strahlentherapie und Onkologie* [Preprint]. Springer Science and Business Media Deutschland GmbH. doi:10.1007/s00066-024-02262-2.
- Feulner, J. *et al.* (2013) ‘Lymph node detection and segmentation in chest CT data using discriminative learning and a spatial prior’, *Medical Image Analysis*, 17(2), pp. 254–270. doi:10.1016/j.media.2012.11.001.

- Finnegan, R.N. *et al.* (2023) 'Open-source, fully-automated hybrid cardiac substructure segmentation: development and optimisation', *Physical and Engineering Sciences in Medicine*, 46(1), pp. 377–393. doi:10.1007/s13246-023-01231-w.
- Fowler, Z. *et al.* (2023) 'Clinical Trial Active Learning', in *Proceedings of the 14th ACM International Conference on Bioinformatics, Computational Biology, and Health Informatics*. New York, NY, USA: ACM, pp. 1–10. doi:10.1145/3584371.3612961.
- Huang, P. *et al.* (2024) 'Prior information guided deep-learning model for tumor bed segmentation in breast cancer radiotherapy', *BMC Medical Imaging*, 24(1). doi:10.1186/s12880-024-01469-0.
- Isensee, F. *et al.* (2021) 'nnU-Net: a self-configuring method for deep learning-based biomedical image segmentation', *Nature Methods*, 18(2), pp. 203–211. doi:10.1038/s41592-020-01008-z.
- Jungo, A. *et al.* (2018) 'On the Effect of Inter-observer Variability for a Reliable Estimation of Uncertainty of Medical Image Segmentation', in, pp. 682–690. doi:10.1007/978-3-030-00928-1_77.
- Kim, D.D. *et al.* (2024) 'Active Learning in Brain Tumor Segmentation with Uncertainty Sampling and Annotation Redundancy Restriction', *Journal of Imaging Informatics in Medicine* [Preprint]. doi:10.1007/s10278-024-01037-6.
- Leong, T. *et al.* (2015) 'TOPGEAR: A randomised phase III trial of perioperative ECF chemotherapy versus preoperative chemoradiation plus perioperative ECF chemotherapy for resectable gastric cancer (an international, intergroup trial of the AGITG/TROG/EORTC/NCIC CTG)', *BMC Cancer*, 15(1). doi:10.1186/s12885-015-1529-x.
- Leong, T. *et al.* (2024) 'Preoperative Chemoradiotherapy for Resectable Gastric Cancer', *New England Journal of Medicine*, 391(19), pp. 1810–1821. doi:10.1056/NEJMoa2405195.
- Leong, Trevor *et al.* (2017) 'TOPGEAR: A Randomized, Phase III Trial of Perioperative ECF Chemotherapy with or Without Preoperative Chemoradiation for Resectable Gastric Cancer: Interim Results from an International, Intergroup Trial of the AGITG, TROG, EORTC and CCTG', *Oncol*, 24, pp. 2252–2258. doi:10.1245/s10434-017-5830-6.
- Liu, Y. *et al.* (2023) 'Imbalanced data classification: Using transfer learning and active sampling', *Engineering Applications of Artificial Intelligence*, 117. doi:10.1016/j.engappai.2022.105621.
- Lukovic, J. *et al.* (2023) 'The Feasibility of Quality Assurance in the TOPGEAR International Phase 3 Clinical Trial of Neoadjuvant Chemoradiation Therapy for Gastric Cancer (an Intergroup Trial of the AGITG/TROG/NHMRC CTC/EORTC/CCTG)', *International Journal of Radiation Oncology Biology Physics*, 117(5), pp. 1096–1106. doi:10.1016/j.ijrobp.2023.06.011.

- Mahapatra, D. *et al.* (2018) 'Efficient active learning for image classification and segmentation using a sample selection and conditional generative adversarial network', in *Lecture Notes in Computer Science (including subseries Lecture Notes in Artificial Intelligence and Lecture Notes in Bioinformatics)*. Springer Verlag, pp. 580–588. doi:10.1007/978-3-030-00934-2_65.
- Maier-Hein, L. *et al.* (2024) 'Metrics reloaded: recommendations for image analysis validation', *Nature Methods*, 21(2), pp. 195–212. doi:10.1038/s41592-023-02151-z.
- Mathai, T.S., Liu, B. and Summers, R.M. (2024) 'Segmentation of mediastinal lymph nodes in CT with anatomical priors', *International Journal of Computer Assisted Radiology and Surgery*, 19(8), pp. 1537–1544. doi:10.1007/s11548-024-03165-4.
- Men, K. *et al.* (2020) 'Automated Quality Assurance of OAR Contouring for Lung Cancer Based on Segmentation With Deep Active Learning', *Frontiers in Oncology*, 10. doi:10.3389/fonc.2020.00986.
- Mir, R. *et al.* (2020) 'Organ at risk delineation for radiation therapy clinical trials: Global Harmonization Group consensus guidelines: GHG OAR consensus contouring guidance', *Radiotherapy and Oncology*, 150, pp. 30–39. doi:10.1016/j.radonc.2020.05.038.
- Nath, V. *et al.* (2021) 'Diminishing Uncertainty within the Training Pool: Active Learning for Medical Image Segmentation', *IEEE Transactions on Medical Imaging*, 40(10), pp. 2534–2547. doi:10.1109/TMI.2020.3048055.
- Nikolov, S. *et al.* (2018) 'Deep learning to achieve clinically applicable segmentation of head and neck anatomy for radiotherapy'. Available at: https://arxiv.org/pdf/1809.04430.pdf (Accessed: 8 April 2019).
- Peters, L.J. *et al.* (2010) 'Critical impact of radiotherapy protocol compliance and quality in the treatment of advanced head and neck cancer: Results from TROG 02.02', *Journal of Clinical Oncology*, 28(18), pp. 2996–3001. doi:10.1200/JCO.2009.27.4498.
- Ren, P. *et al.* (2022) 'A Survey of Deep Active Learning', *ACM Computing Surveys*. Association for Computing Machinery. doi:10.1145/3472291.
- Sahiner, B. *et al.* (2019) 'Deep learning in medical imaging and radiation therapy', *Medical Physics*, pp. e1–e36. doi:10.1002/mp.13264.
- Sidhom, M.A. *et al.* (2008) 'Post-prostatectomy radiation therapy: Consensus guidelines of the Australian and New Zealand Radiation Oncology Genito-Urinary Group', *Radiotherapy and Oncology*, 88(1), pp. 10–19. doi:10.1016/j.radonc.2008.05.006.
- Sorensen, T.A. (1948) 'A method of establishing groups of equal amplitude in plant sociology based on similarity of species content and its application to analyses of the vegetation on Danish commons', *Biol. Skar.*, 5, pp. 1–34.

- Tajbakhsh, N. *et al.* (2020) ‘Embracing imperfect datasets: A review of deep learning solutions for medical image segmentation’, *Medical Image Analysis*, 63. doi:10.1016/j.media.2020.101693.
- Wang, H. *et al.* (2024) ‘A comprehensive survey on deep active learning in medical image analysis’, *Medical Image Analysis*. Elsevier B.V. doi:10.1016/j.media.2024.103201.
- Wang, K. *et al.* (2023) ‘Improving brain tumor segmentation with anatomical prior-informed pre-training’, *Frontiers in Medicine*, 10. doi:10.3389/fmed.2023.1211800.
- Warfield, S.K., Zou, K.H. and Wells, W.M. (2004) ‘Simultaneous truth and performance level estimation (STAPLE): An algorithm for the validation of image segmentation’, *IEEE Transactions on Medical Imaging*, 23(7), pp. 903–921. doi:10.1109/TMI.2004.828354.
- Wasserthal, J. *et al.* (2023) ‘TotalSegmentator: Robust Segmentation of 104 Anatomic Structures in CT Images’, *Radiology: Artificial Intelligence*, 5(5). doi:10.1148/ryai.230024.
- Xu, L. *et al.* (2021) ‘Clinical target volume segmentation for stomach cancer by stochastic width deep neural network’, *Medical Physics*, 48(4), pp. 1720–1730. doi:10.1002/mp.14733.